# Many-body hybrid Excitons in Organic-Inorganic van der Waals Heterostructures

Shaohua Fu,[1,2,4] [#] Jianwei Ding[3] [#], Haifeng Lv,[5] Shuangyan Liu,[1] Kun Zhao[1], Zhiying Bai[1], Dawei He,[1] Rui Wang,[6] Jimin Zhao,[4] Xiaojun Wu,[5] Dongsheng Tang,[2] [*] Xiaohui Qiu,[3] [*] Yongsheng Wang[1], Xiaoxian Zhang[1] [*]

[1]*Key Laboratory of Luminescence and Optical Information, Ministry of Education, Institute of Optoelectronic Technology, Beijing Jiaotong University, Beijing 100044, P. R. China*

[2]*Synergetic Innovation Center for Quantum Effects an Application, Key Laboratory of Low-dimensional Quantum Structures and Quantum Control of Ministry of Education, School of Physics and Electronics, Hunan Normal University, Changsha 410081, China*

[3]*CAS Key Laboratory of Standardization and Measurement for Nanotechnology, CAS Center for Excellence in Nanoscience, National Center for Nanoscience and Technology, Beijing 100190, P. R. China. University of Chinese Academy of Science, Beijing 100049, P. R. China.*

[4]*Beijing National Laboratory for Condensed Matter Physics, Institute of Physics, Chinese Academy of Sciences, Beijing 100190, P. R. China*

[5]*Hefei National Laboratory for Physical Sciences at the Microscale, CAS Key Laboratory of Materials for Energy Conversion, Synergetic Innovation of Quantum Information & Quantum Technology, School of Chemistry and Materials Sciences, and CAS Center for Excellence in Nanoscience, University of Science and Technology of China, Hefei, Anhui 230026, P.R. China*

[6]*Beijing Information technology college, Beijing 100015, P. R. China*

#These authors contributed equally

*e-mail: dstang@hunnu.edu.cn; xhqiu@nanoctr.cn; zhxiaoxian@bjtu.edu.cn

**Keywords:** hybrid exciton, organic-inorganic heterostructure, low temperature photoluminescence, molecular orientation, interfacial coupling

# Abstract

The emerging two-dimensional material (2D)–organic heterostructures are promising candidates for novel optoelectronic applications due to their rich excitonic physics. The hybridization at the organic-inorganic interface can give rise to intriguing hybrid excitons that combine the advantages of the Wannier-Mott and Frenkel excitons simultaneously, featuring a momentum-direct nature. Here, we report hybrid excitons at the copper phthalocyanine/molybdenum diselenide (CuPc/MoSe$_2$) interface with abnormal temperature dependence using low-temperature photoluminescence spectroscopy. The new emission peaks observed in the CuPc/MoSe$_2$ heterostructure indicate the formation of interfacial hybrid excitons. The density functional theory (DFT) calculation confirms the strong hybridization between the lowest unoccupied molecular orbital (LUMO) of CuPc and the conduction band minimum (CBM) of MoSe$_2$, suggesting that the hybrid excitons consist of electrons extended in both layers and holes confined in individual layers. The temperature-dependent behaviors of hybrid excitons show that they gain the signatures of the Frenkel excitons of CuPc and the Wannier-Mott excitons of MoSe$_2$ simultaneously. The out-of-plane molecular orientation is further used to tailor the interfacial hybrid exciton states. Our results reveal the hybrid excitons at the CuPc/MoSe$_2$ interface with tunability by molecular orientation, which suggests that the emerging organic-inorganic heterostructure can be a promising platform for many-body exciton physics.

# Introduction

The rich excitonic physics found in van der Waals heterostructures provides new platform for novel optoelectronics and future quantum technologies[1-8]. Hybrid excitons are many-body exciton states originated from the hybridization of electronic states at interfaces[9-14], displaying potential in quantum optics[15] and strongly correlated electronic physics[10]. The hybridization at the organic-inorganic interface could lead to intriguing new hybrid Wannier-Mott-Frenkel excitons[16,17], which are proposed to gain the advantages of the Wannier-Mott excitons and Frenkel simultaneously, showing potential in nonlinear optics. Nevertheless, the coupling at the organic-inorganic interfaces is generally weak[18], and the direct evidence of the hybrid excitons at organic-inorganic interface is still elusive.

Transition metal dichalcogenides (TMDs) are promising building blocks for realizing such hybrid excitons, because they not only contain rich exciton physics[19-26], but can also serve as a suitable template for the growth of well-ordered organic films [27,28]. To achieve the new hybrid Wannier-Mott-Frenkel excitons, interfacial hybridization should be established between the discrete organic electronic states and the continuous inorganic electronic states. The theoretical calculation suggests that there could exist strong interfacial hybridization between the LUMO of organics and the CBM of the TMDs[29]. Moreover, the short-range interactions, such as ultrafast charge transfer[30-33] has been observed at the organic-TMD interfaces, suggesting that the coherent superposition of the electron wavefunctions and the formation of hybrid excitons are possible at such interfaces. Besides, the excitons formed at the organic-inorganic heterostructure interface are predicted to have a momentum-direct nature[29], which could simplify the fabrication process and maintain the novel exciton physics at the same time.

It is well-known that the interfacial hybridization strength depends sensitively on the interlayer twist angle and stacking configuration in inorganic heterostructures[12,13]. Similarly, the molecular orientation at the organic-inorganic interface should be capable of tuning the interfacial distance and atomic stacking, thus the interfacial hybridization strength and the hybrid exciton behavior could be effectively tailored, which would be beneficial for designing the organic-inorganic interface functionality in the future.

In this article, we report the formation of hybrid excitons at the $CuPc/MoSe_2$ interface and their further modulation by molecular orientation. Combination of the new emission peaks observed in photoluminescence spectra and the further DFT calculations confirm the emergence of the new hybrid excitons. Meanwhile, the temperature-dependent measurements reveal that the hybrid excitons combine the signature of both Wannier-Mott and Frenkel exciton species. Furthermore, the out-of-plane molecular orientation is found to effectively tailor the interfacial hybrid excitons. Our results suggest that the organic-inorganic heterostructure is a promising platform to explore many-body exciton physics such as exciton condensates[29], and further optoelectrical applications including light harvesting and photodetection[28].

# Results and Discussion

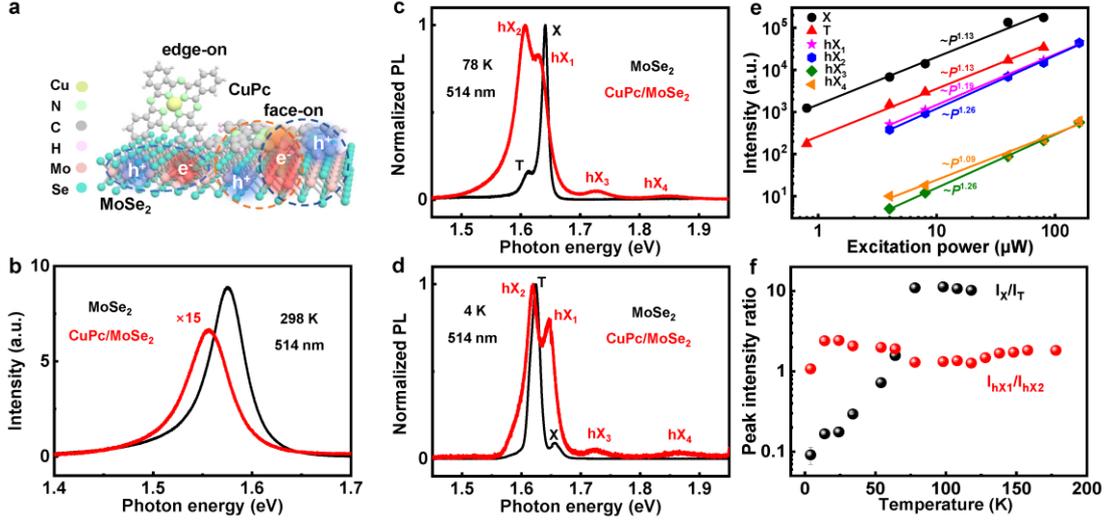

**Figure 1.** Sample configuration and optical characterization. **a** Schematic illustration of the different molecular orientations (face-on and edge-on) at CuPc/MoSe$_2$ heterostructure interface. **b** PL spectra of MoSe$_2$ and CuPc/MoSe$_2$ heterostructure at 298 K. **c** PL spectra of the MoSe$_2$ and CuPc/MoSe$_2$ heterostructure at 78 K. **d** PL spectra of the MoSe$_2$ and CuPc/MoSe$_2$ heterostructure at 4 K. **e** The PL peak intensity of MoSe$_2$ and CuPc/MoSe$_2$ heterostructure as a function of the excitation power. **f** The PL peak intensity ratio of exciton versus trion in MoSe$_2$ ($I_X/I_T$) and hX$_1$ versus hX$_2$ in the CuPc/MoSe$_2$ heterostructure ($I_{hX1}/I_{hX2}$) as a function of temperature.

Figure 1a shows a schematic of the CuPc/MoSe$_2$ heterostructure configuration. We consider two different molecular orientations, *i.e.,* face-on and edge-on, at the CuPc/MoSe$_2$ interface, which will sensitively influence the interfacial coupling strength and exciton behaviors. The CuPc/MoSe$_2$ heterostructures are prepared by directly evaporating CuPc molecules on top of monolayer MoSe$_2$ surface in vacuum (see details in methods). A film thickness of ~5 nm is determined by AFM (Figure S1b). The optimal molecular orientation at interface of the as-grown sample is the face-on orientation, which has been reported in similar systems[33,34] and revealed by our theoretical calculation later. The edge-on orientation is introduced by using the CuPc single crystal later. In the face-on orientation, the planar conjugated structure of the CuPc molecule[35] and the atomic flat surface of monolayer MoSe$_2$[36] without dangling bonds facilitate interfacial coupling between them (Figure 1a). The photoluminescence (PL) spectra of MoSe$_2$ and CuPc/MoSe$_2$ heterostructure acquired at room temperature are shown in Figure 1b and Figure S1e. The MoSe$_2$ exhibits a pronounced PL peak located at ~1.58 eV originated from the A excitonic transition[37]. In contrast, a remarkable redshift of ~20 m eV in the PL peak energy and a strong quenching in the PL peak intensity are observed in CuPc/MoSe$_2$ heterostructure. The pure CuPc thin film shows no detectable PL signal (Figure S2) due to the weak absorption at approximately 514 nm[33] and its strong intersystem crossing[38]. The thickness of CuPc thin film is found to have negligible influence on the PL quenching ratio of heterostructure (Figure S1f and Figure S3), and the dielectric environment change [39,40] also has little effect on the phenomenon in another heterostructure with opposite configuration (Figure S4), indicating that the observed phenomenon should originate from the intrinsic interfacial coupling between CuPc and MoSe$_2$. Time-resolved PL measurements show that the heterostructure exhibits much shorter PL lifetime than MoSe$_2$ (Figure S5), suggesting that the redshift behavior in the PL of heterostructure should not originate from charge transfer exciton[41].

Low-temperature PL spectra under 514 nm excitation are obtained at 78 K to further reveal the possible mechanism. We still observe no detectable PL signal in the pure CuPc thin film (Figure S6a). As displayed in Figure 1c, two emission peaks located at ~1.648 eV and 1.618 eV are observed in the PL spectrum of MoSe$_2$, which can be ascribed to the emission from the A exciton (X) and trion (T) of MoSe$_2$[37]. A striking contrast is observed in the PL spectrum of heterostructure, with four new emission peaks located at ~1.630 eV, ~1.606 eV, ~1.727 eV and ~1.848 eV emerging, labeled as hX$_1$, hX$_2$, hX$_3$, and hX$_4$, respectively. The hX$_1$ and hX$_2$ show a clear redshift compared with the A exciton of MoSe$_2$, and the hX$_4$ peak displays an obvious redshift with respect to the B exciton of MoSe$_2$ (Figure S7).

The hX$_3$ peak is a totally new PL peak that is not observed in pure monolayer MoSe$_2$ and CuPc thin films. When we resonantly excite CuPc thin film using 633 nm laser, the hX$_3$ peak is still different from the weak and broad PL signal of CuPc thin film (Figure S6b-d). To avoid the influence the CuPc thin film morphology, another heterostructure is prepared via dry transfer method to keep the same morphology, whereas we can still observe the hX$_3$ peak even without annealing (Figure S8a-b), suggesting that the hX$_3$ peak should originate from the interface of heterostructure. The general case of new emission peak at organic-inorganic interface is the charge transfer exciton[41,42], but the hX$_3$ peak has a much higher energy (~79 m eV) than the A exciton of MoSe$_2$, so charge transfer exciton may not be a reasonable explanation here. The dark exciton[43] of MoSe$_2$ is also excluded due to the much larger energy of hX$_3$.

Power-dependent PL spectra are further obtained to examine the origin of the new peaks in heterostructure (Figure 1e and Figure S8c-d). The relationship between excitation power and PL intensity can be expressed as[44] $I \propto P^\alpha$, in which $I$ represents the PL intensity and $P$ represents the excitation power. The intensities of X and T peaks in MoSe$_2$ show a linear relationship with excitation power with a slope of ~1.13, which indicates recombination from excitons[45]. Interestingly, all the new peaks in heterostructure also slow the linear relationship with similar slopes, which suggests similar exciton behavior with no biexciton[46] or defect effect[47]. In addition, it's clear that the PL peaks in heterostructure show significant broadening compared with those of MoSe$_2$. Since the PL peak redshift and broadening are the signature of interfacial hybridization as reported in similar MoSe$_2$/WS$_2$ heterostructure[12], and we also exclude other possibilities in the above discussions, we propose that the new emission peaks of CuPc/MoSe$_2$ heterostructure may be hybrid excitons due to interfacial hybridization.

The energy difference between hX$_1$ and hX$_2$ peak is close to the trion binding energy of MoSe$_2$[37], so we assign the two peaks to the hybridized A exciton and hybridized trion, respectively. We further perform PL measurements at 4 K to examine the influence of interfacial hybridization. As illustrated in Figure 1d, the PL spectrum of MoSe$_2$ at 4 K is dominated by trion rather than A exciton due to the enhanced trion localization, in great contrast to that at 78 K[48]. On the contrary, the PL spectrum of heterostructure is still dominated by hX$_1$ and hX$_2$, similar to that at 78 K. The evolution of the PL of MoSe$_2$ from 4 K to 200 K clear shows that the intensity ratio of X/T has increased from 0.09 to 10 when the temperature is increased from 4 K to 200 K due to the increased thermal perturbance to trion formation (Figure 1f and Figure S9a). Nevertheless, the intensity radio of hX$_1$/hX$_2$ shows a weak temperature dependence from 4 K to 200 K (Figure 1f and Figure S9c), differing from the pure exciton and trion behavior in MoSe$_2$, which indicates that the interfacial hybridization effect has changed the trion localization condition. In addition, the PL spectrum of heterostructure displays a highly asymmetric line-shape with prominent low energy tail (Figure 1c and 1d), which can be ascribed to an energy shakeup process during the recombination of hybrid excitons[11].

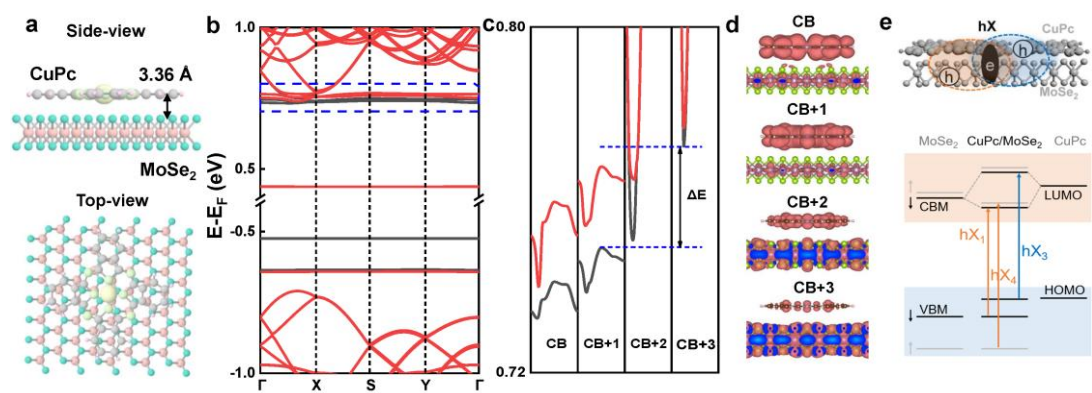

**Figure 2.** Theoretical calculation of electronic structure in CuPc/MoSe$_2$ heterostructure. **a** Top-view and side-view for the optimized structure of CuPc/MoSe$_2$ heterostructure. **b** Calculated band structure of CuPc/MoSe$_2$ heterostructure. **c** Conduction bands (CB), CB+1, CB+2 and CB+3 in the energy range of 0.72 to 0.80 eV, which correspond to the blue square in **b**. **d** Projected charge density for bands in **b**. ΔE (23 meV) is defined as the energy difference between CB+1 and CB+3, which is mostly contributed by CuPc and MoSe$_2$, respectively. **e** Schematic illustration of the formation of interfacial hybrid excitons due to the hybridization between LUMO of CuPc and CBM of MoSe$_2$.

The above results indicate that the interfacial hybridization effect could lead to the formation of hybrid excitons and change the exciton behavior in CuPc/MoSe$_2$ heterostructure. First-principles calculations are further performed to confirm this. The heterostructure is built by adsorbing a CuPc molecule on a 5×3√3 supercell of MoSe$_2$ and the optimal molecular orientation is the face-on orientation (Figure 2a). The calculated electronic structure of CuPc/MoSe$_2$ heterostructure with face-on orientation is shown in Figure 2b. We could recognize two nearly flat bands near 0.5 and -0.5 eV, which correspond to the singly occupied and unoccupied molecular orbitals (SOMO and SUMO) of the CuPc molecule. Then, we concentrate on the bands in the energy range of 0.72 to 0.80 eV (Figure 2c), which are denoted as conduction band CB, CB+1, CB+2 and CB+3. As shown in Figure 2d, the projected charge density shows that CB and CB+1 are mostly contributed by CuPc, and CB+3 is mostly contributed by $MoSe_2$. Notably, CB+2 is contributed both by CuPc and $MoSe_2$, which could be regarded as the hybridization between the LUMO of CuPc and the conduction band of $MoSe_2$. The energy difference between CB+3 and CB+1 in the same spin channel is approximately 23 m eV, leading to strong hybridization between CuPc and $MoSe_2$ at the face-on orientation, which can explain the observed new hybrid excitons at CuPc/$MoSe_2$ interface. The calculation also reveals that the formed hybrid excitons consist of electrons extended in both layers and holes confined in individual layers (Figure 2e), which can be used to achieve novel quantum control at organic-inorganic interfaces[13].

The temperature-dependent behaviors of the observed hybrid excitons are carefully examined from 78 K to 298 K. For CuPc /$MoSe_2$ heterostructure (Figure 3a), we clearly observe a remarkable increase in the whole PL intensity when cooling from room temperature (298 K) to low temperature (78 K), which can be explained by the suppression of nonradiative recombination[49]. When the temperature is higher than 178 K, $hX_2$ becomes undistinguishable and $hX_1$ dominates the PL spectra (Figure 3b). Whereas for $MoSe_2$, the trion peak disappears at 98 K and the A exciton becomes dominant (Figure S10). To our surprise, the peak energy of the hybrid excitons shows different temperature dependence (Figure 3a-c).

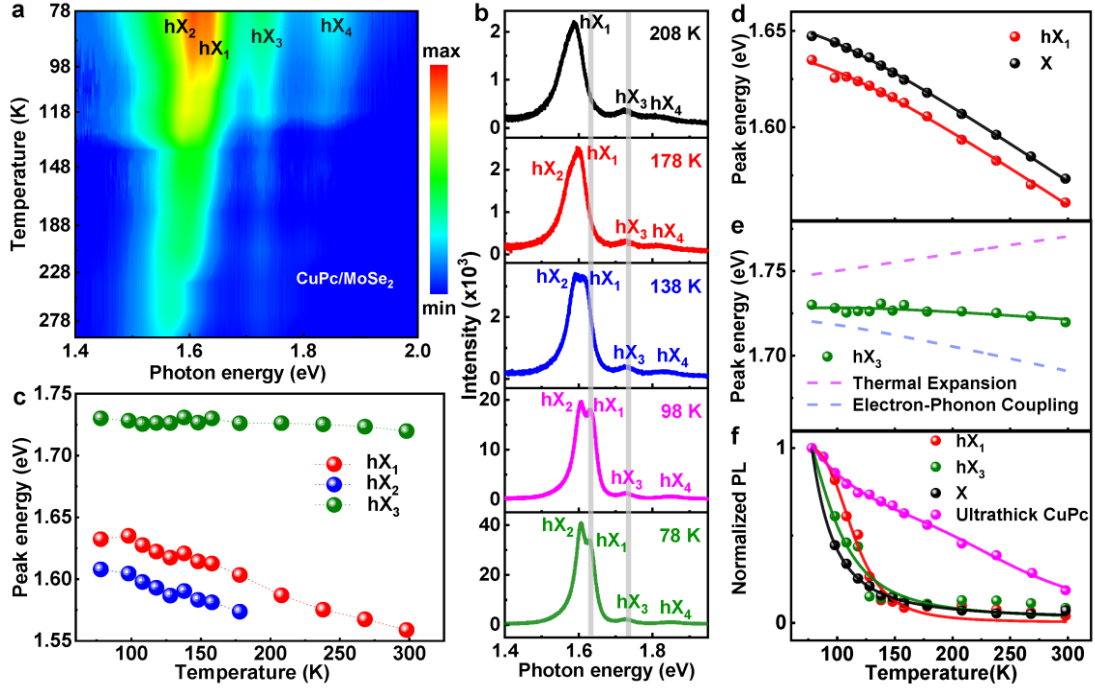

**Figure 3.** Temperature dependence of the hybrid excitons. **a** Two-dimensional PL spectrum of CuPc/MoSe$_2$ heterostructure as a function of temperature. **b** PL spectra of CuPc/MoSe$_2$ heterostructure in the energy range of 1.4 - 1.95 eV at the temperatures of 78 K, 98 K, 138 K, 178 K, and 208 K, respectively. **c** The peak energy of hX$_1$ (black), hX$_2$ (red), and hX$_3$ (purple) as a function of temperature. **d** Temperature dependence and the fitting of the peak energy of X peak of monolayer MoSe$_2$ and hX$_1$ peak of CuPc/MoSe$_2$ heterostructure. **e** Temperature dependence and the fitting of the peak energy of hX$_3$ peak of CuPc/MoSe$_2$ heterostructure, the blue and yellow dotted line represents the contribution of thermal expansion and electron-phonon coupling on the peak energy of hX$_3$. **f** Temperature dependence and the fitting of the peak intensity of X, hX$_1$, hX$_3$ peak and the photoluminescence peak of ultrathick CuPc film.

We firstly focus on the hX$_1$, hX$_2$, and hX$_4$ peaks, of which the peak energy shows obvious redshift with increasing temperature due to the increased electron-phonon interactions[49], similar to the temperature-dependent behavior of exciton and trion in MoSe$_2$ (Figure S10). By fitting the peak energy with the semiconductor bandgap model[50]:

$$E_g(T) = E_g(0) - S\hbar\omega \left[\coth\left(\frac{\hbar\omega}{2k_BT}\right) - 1\right] \qquad (1)$$

where $E_g$ represents the bandgap, $\hbar\omega$ represents the phonon energy, $S$ represents the electron-phonon coupling strength, and $T$ represents the temperature, we obtain a similar phonon energy for the CuPc/MoSe$_2$ heterostructure and MoSe$_2$ (Figure 3d, and Table S1), suggesting that these hybrid excitons are strongly influenced by the phonons of MoSe$_2$. Besides, the obtained bandgap of MoSe$_2$ is decreased in heterostructure due to the effect of interfacial hybridization (Table S1).

The hX$_3$ peak located at ~1.72 eV is more unique among the four hybrid excitons. The peak energy shows a rather weak temperature dependence (Fig 3a-c), which is totally different from the other three peaks. Such weak temperature-dependent behavior of excitons has been observed in organic molecules, in which the effects of thermal expansion and exciton-phonon coupling almost cancel out[51]. The peak energy of hX$_3$ peak is well fitted using the model considering the thermal expansion and electron-phonon coupling effect simultaneously[52]:

$$E_g(T) = E_g(0) + A_{TE}T + A_{EP}\left[\frac{2}{\exp\left(\frac{\hbar\omega}{k_BT}\right)-1}+1\right] \quad (2)$$

where $A_{TE}$ is the weighting factor of thermal expansion effect, $A_{EP}$ represents the weighting factor of electron-phonon coupling. According to the fitting results (Fig 3e and Table S2), it's evident that the blueshift caused by thermal expansion and the redshift resulted from electron-phonon coupling almost cancel out, leading to the abnormal weak temperature-dependent behavior of hX$_3$. We also observe similar weak temperature-dependent behavior in an ultrathick CuPc film (Figure S11), suggesting that the hX$_3$ peak displays the signature of Frenkel excitons of CuPc.

However, the hX$_3$ peak cannot be assigned to the emission of pure CuPc molecules as discussed above (Figure S6). More importantly, the hX$_3$ peak also combines the character of the Wannier-Mott exciton of MoSe$_2$ as its peak intensity shows similar temperature-dependent behavior to the A exciton of MoSe$_2$ (Figure 3f), different from the behavior of ultrathick CuPc film. The peak intensity is fitted using the excitation energy model of photoluminescence intensity[53]:

$$I(T) = I_0/[1 + \exp\left(\frac{E_1}{k_B T}\right)] \tag{3}$$

where $I(T)$ represents the PL intensity at temperature $T$, $E_1$ represents the excitation energy, and the corresponding fitting results are shown in Fig 3e and Table S2. The obtained excitation energy of ~ 30 meV of the A exciton of MoSe$_2$ is similar to the splitting energy of the conduction band[54], suggesting that the bright A excitons are transformed into dark A excitons with increasing of temperature. To our surprise, the hX$_3$ displays similar excitation energy to the A exciton of MoSe$_2$, which indicates that it gains the character of the Wannier-Mott exciton of MoSe$_2$ due to interfacial hybridization. Therefore, the peak energy of hX$_3$ shows the signature of the Frenkel excitons in the organic CuPc thin film and its peak intensity display the character of Wannier-Mott excitons of the inorganic MoSe$_2$ monolayer, which unambiguously reveal the formation of hybrid Frenkel-Wannier-Mott excitons at the CuPc/MoSe$_2$ interface.

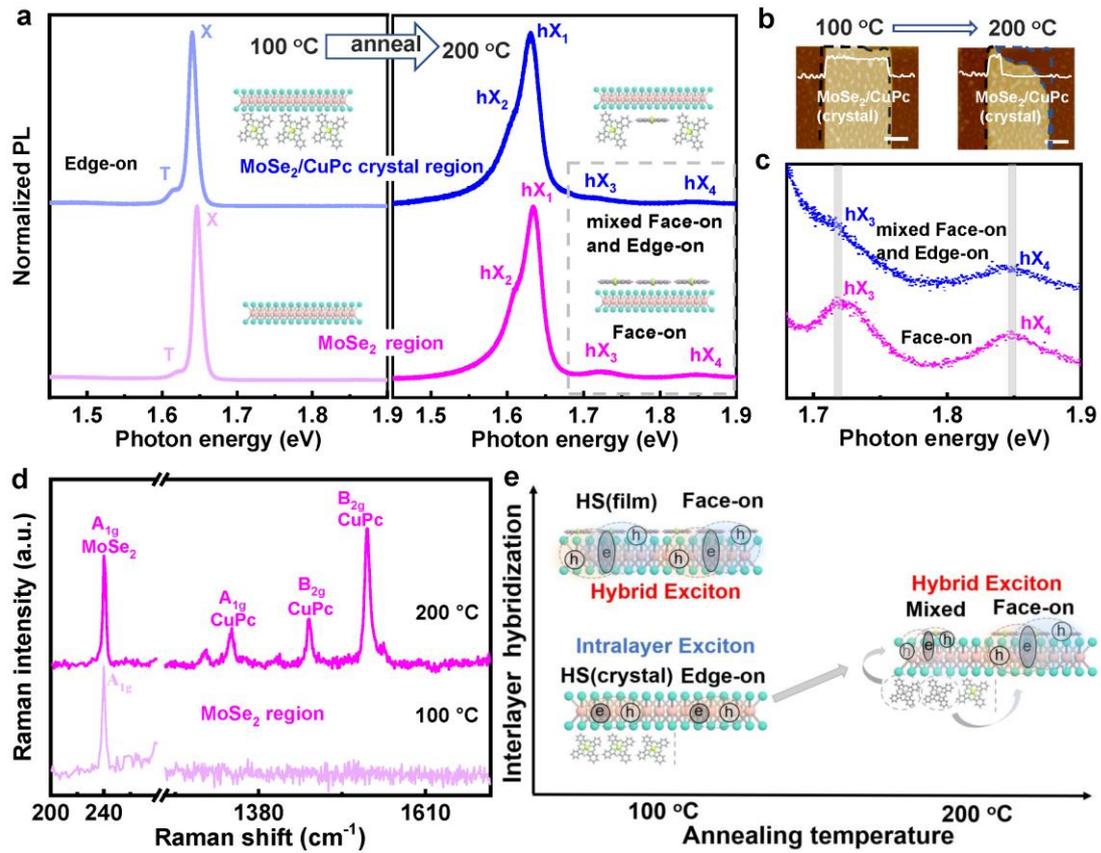

**Figure 4.** Molecular orientation-dependent hybrid excitons. **a** The PL spectra in the $MoSe_2$ region and $MoSe_2$/CuPc crystal region for the same $MoSe_2$/CuPc crystal heterostructure after annealing at 100°C and 200°C. After annealing at 200°C, the CuPc crystal partially decomposes and the CuPc molecules can migrate on $MoSe_2$, which lead to the formation hybrid excitons in both regions. **b** AFM topographic image of the $MoSe_2$/CuPc crystal heterostructure after annealing at 100°C and 200°C. **c** Enlarged view of the image in the gray dotted box in **a**. **d** Raman spectra of the $MoSe_2$ region in the $MoSe_2$/CuPc crystal heterostructure sample after annealing at 100°C and 200°C. **e** Schematic illustration of the relationship between molecular orientation (face-on, edge-on) of CuPc and interlayer hybridization.

The molecular orientation is introduced as a new degree of freedom to modulate the interfacial hybridization strength, which will further tailor the hybrid excitons. In general, the CuPc molecule tends to adopt a face-on orientation on the $MoSe_2$ surface that allows efficient interfacial hybridization, as revealed by our theoretical calculation. In contrast, the edge-on orientation will experience insufficient interfacial hybridization due to the larger interfacial distance. To demonstrate that the molecular orientation can be used to tailor the interfacial hybrid exciton states, we carefully prepared $MoSe_2$/CuPc film heterostructure and $MoSe_2$/CuPc crystal heterostructure simultaneously by the dry transfer method. Since the CuPc molecule stacks randomly in the CuPc film under our preparation condition[55], it easily adopts a face-on orientation on the $MoSe_2$ surface. However, because the CuPc molecule shows a herringbone stacking in the crystal[56], it can only adopt an edge-on orientation on $MoSe_2$ surface before crystal decomposition (Figure S12c). Because the monolayer $MoSe_2$ partially covers the CuPc crystal, we could compare the measurements from the $MoSe_2$ region and $MoSe_2$/CuPc crystal region in the same sample (Figure S12c). No obvious quenching and hybrid excitons are observed in the crystal heterostructure (Figure S12c-d). After annealing simultaneously at 100 °C, the PL spectra of $MoSe_2$/CuPc film heterostructure and $MoSe_2$/CuPc crystal heterostructure display great contrast as expected. The PL of crystal heterostructure shows similar spectral features and slight quenching compared with monolayer $MoSe_2$ (Figure 4a and Figure S13a), indicating a weak interfacial hybridization strength. However, the PL of film heterostructure presents obvious quenching and the formation of hybrid excitons (Figure S13a-b). The influence of the morphology of CuPc can be excluded since the surface of CuPc crystal is flatter than that of the CuPc film (Figure S12a). Therefore, the above results suggest that the molecular orientation can be used to tune the interlayer hybridization strength and further tailor the interfacial hybrid excitons.

To confirm the above deduction, the MoSe$_2$/CuPc crystal heterostructure is further annealed at 200 °C in vacuum to decompose the CuPc crystal. When the CuPc crystal decomposes, the CuPc molecules can easily adopt a face-on orientation on the MoSe$_2$ surface, thus, the interfacial hybrid excitons should also be observed. After annealing at 200 °C, the PL spectra display obvious quenching in both the MoSe$_2$/CuPc crystal region and MoSe$_2$ region (Figure S13c), and clearly shows the formation of hybrid excitons (Figure 4a and 4c), which indicates that the molecular orientation is changed from edge-on to face-on after CuPc crystal decomposition. The decomposition of the CuPc crystal is confirmed by AFM, as shown in Figure 4b and Figure S14. It is obvious that the CuPc crystal partially decomposes after annealing at 200 °C, as evidenced by the change in the AFM height profile. Note that we can even observe hybrid excitons in the MoSe$_2$ region because the CuPc molecules can migrate on MoSe$_2$ surface after the decomposition of CuPc crystal. The Raman spectra further confirms this, as the Raman peaks of both MoSe$_2$ and CuPc molecules appear in the MoSe$_2$ region after annealing at 200 °C (Figure 4d). These results unambiguously show that we can successfully tailor the interfacial hybrid excitons by changing the molecular orientation (Figure 4e). The theoretical calculation also supports our results. As shown in the electronic structure of heterostructure at the edge-on orientation (Figure S15), we observe no obvious interfacial hybridization, which coincides with the observed phenomena in MoSe$_2$/CuPc crystal heterostructure.

## Conclusion

We have demonstrated the formation of new hybrid Wannier-Mott-Frenkel excitons in CuPc/MoSe$_2$ heterostructure due to the interfacial hybridization between the LUMO of CuPc and the CBM of MoSe$_2$, which is further rationalized by the first principles calculations. The new hybrid excitons consist of electrons delocalized in both layers and holes confined in individual layer, enabling simultaneous large optical and electrical dipoles[13]. The peak energy of hX$_3$ shows an abnormal weak temperature dependence similar to the organic Frenkel excitons, which is explained by the combination of the thermal expansion and electron-phonon coupling effect. Whereas the peak intensity of hX$_3$ displays similar non-radiative recombination excitation energy to MoSe$_2$, suggesting that the hybrid excitons simultaneously gain the signature of the Wannier-Mott excitons in MoSe$_2$ and the Frenkel excitons in CuPc. The excellent agreement between the theoretical and experimental results not only validates the observed intriguing many-body exciton phenomenon, but also provides a basis for manipulating hybrid excitons at the organic-inorganic interface. For instance, the large electrical dipole in the out-of-plane direction can be used to achieve novel electrical control of the hybrid excitons[57,58]. Our result is also of great importance for realizing tunable interlayer hybridization strength by changing the molecular orientation, which can be used to tailor the exciton states at the organic-inorganic interface. In conclusion, we report the formation of hybrid Wannier-Mott-Frenkel excitons with strong molecular orientation dependence and abnormal temperature-dependent behavior that originate from interfacial hybridization, which is meaningful for the many-body exciton physics such as exciton condensation and potential optoelectrical applications at the organic-inorganic interface.

## Experimental Section

**Sample preparation**

**(1) Construction of the CuPc film /MoSe$_2$ heterostructure**

Monolayer MoSe$_2$ was mechanically exfoliated on a SiO$_2$/Si substrate from bulk crystals and further annealed in vacuum at 200 °C to remove surface contaminants. The thickness of MoSe$_2$ was confirmed by optical contrast, atomic force microscopy (AFM), and Raman measurements (Figure S1). To construct the CuPc (film)/MoSe$_2$ heterostructure, CuPc thin film was directly deposited on top of monolayer MoSe$_2$ using thermal evaporation in vacuum (home-built evaporator). The heating current was maintained at 5 amperes and the average evaporation speed was 0.25 nm/min.

**(2) Construction of the MoSe$_2$/CuPc film heterostructure**

The CuPc film was firstly thermally evaporated on a SiO$_2$/Si substrate using the same conditions as (1). Then, monolayer MoSe$_2$ was mechanically exfoliated on the PDMS substrate from bulk crystals, and further transferred on top of the CuPc film using the dry transfer method.

**(3) Construction of the MoSe$_2$/CuPc crystal heterostructure**

The single crystals of CuPc were grown by the physical vapor deposition (PVT) method in a quartz tube with a hot zone temperature of 400°C. To construct the MoSe$_2$/CuPc (crystal) heterostructure, monolayer MoSe$_2$ was mechanically exfoliated on a PDMS substrate, and further transferred on top of a CuPc single crystal using the dry transfer method.

**Low temperature PL Measurements.**

The measurements at 78 K were conducted in a temperature-controlled cryostat (THMS600, Linkam) with a diffraction-limited excitation beam diameter of 1μm. The signal was collected using a 50X long-working distance objective and detected on a commercial Renishaw inVia spectrometer. The excitation power was selected to be below 200 μW to avoid heating damage to the sample. The measurements at 4 K were conducted in a temperature-controlled cryostat (Montana Instruments) with an excitation beam diameter of 1μm. The signal was collected using a 100X objective and detected on a commercial Ocean Optics spectrometer.

**Theoretical calculation.** First-principles calculations were carried out based on the density functional theory (DFT) framework by utilizing the Vienna Ab initio Simulation Package (VASP) 5.4.4 package[59,60]. Pseudopotentials were used to describe the electron-ion interactions within the PAW approach and generalized gradient approximations (GGA) of Perdew-Burke-Ernzerhof (PBE) were adopted for the exchange-correlation potential[61-63]. To better describe the interlayer van der Waals (vdW) interactions, we adopt optB88-vdW corrections for the optimization of structures[64]. The electron wave functions are expanded on a plane-wave basis set with an energy cutoff of 520 eV. The atomic coordinates of all structures were allowed to relax until the forces acting on the ions were less than 0.01 eV Å$^{-1}$. The convergence criterion for the electronic self-consistent cycle is fixed at $1\times10^{-5}$ eV. The integrations in the reduced Brillouin zone are performed on a 3×3×1 Monkhorst-Pack special k-points for optimization and self-consistent calculations[65,66]. A vacuum slab above 15 Å was used in all calculations to avoid interlayer interactions. The CuPc/MoSe$_2$ heterostructure is modeled by adsorbing one CuPc molecule on a 5×3√3 supercell of MoSe$_2$, which can be written as Mo30Se60C32N8H16Cu. The lattice parameters of the CuPc/MoSe$_2$ heterostructure were calculated to be a = 16.62 Å, b = 17.27 Å, and α=β=γ=90°. The interlayer distance between CuPc and MoSe$_2$ substrate is approximately 3.36 Å.×

## Supporting Information

Supporting Information is available from the Wiley Online Library or from the author.

## Acknowledgements

This work was supported by the National Nature Science Foundation of China (Grant Nos. 11974088, 12074116, 21790353, 61875236, 61975007), the National Key Research and Development Program of China (Grant Nos.2016YFA0202302, 2017YFA0205000, 2021YFA1400201), the Strategic Priority Research Program of CAS (Grant No. XDB30000000), the CAS Project for Young Scientists in Basic Research (Grant No. YSBR-059).

## Conflict of Interest

The authors declare no conflict of interest.

# Author Contributions

X.Z., X.Q. and D.T. conceived the idea; S.F. and J.D. prepared the samples and conducted all the optical measurements and the corresponding data analysis; X.W. and H.L. performed the DFT calculations; This manuscript was prepared primarily by X.Z., S.F. and J.D., and all authors contributed to discussing and commenting on the paper.

# Data Availability Statement

The data that support the findings of this study are available from the corresponding authors upon reasonable request.